\documentclass[12pt,a4papper]{article}
\usepackage{graphics}
\usepackage{graphicx}
\usepackage{amssymb}

\begin{document}

\title{Time, clocks, parametric invariance and the Pioneer Anomaly v2}
\author{Antonio F. Ra\~nada\thanks{Facultad de
F\'{\i}sica, Universidad Complutense, 28040 Madrid, Spain. E-mail
afr@fis.ucm.es, corresponding author.} \and Alfredo
Tiemblo\thanks{Instituto de Matem\'aticas y F\'{\i}sica Fundamental,
Consejo Superior de Investigaciones Cient\'ificas, Serrano 113b,
28006 Madrid, Spain.}}
\date{23 October 2006}
\maketitle

\begin{abstract}

In the context of a parametric theory (with the time being a
dynamical variable) we consider the coupling between the quantum
vacuum and the background gravitation that pervades the universe
(unavoidable because of the universality of gravity). In our model
the fourth Heisenberg relation introduces a possible source of
discrepancy between the marches of atomic and gravitational clocks,
which accelerate with respect to one another. This produces another
discrepancy between the observations, performed with atomic time,
and the theoretical analysis, which uses parametric astronomical
time. Curiously, this approach turns out to be compatible with
current physics; lacking a unified theory of quantum physics and
gravitation, it cannot be discarded {\it a priori}. It turns out
that this phenomenon has the same footprint as the Pioneer Anomaly,
what suggests a solution to this riddle. This is because the
velocity of the Pioneer spaceship with respect to atomic time turns
out to be slightly smaller that with astronomical time, so that that
the apparent trajectory lags behind the real one. In 1998, after
many unsuccessful efforts to account for this phenomenon, the
discoverers suggested ``the possibility that the origin of the
anomalous signal is new physics".

\end{abstract}

\newpage

\section{Introduction}

 The problem of time is one of
the most obscure and controversial in the history of knowledge,
and has obvious implications in all the fields of thought.
Sticking to physics, the problem is initially posed in the
dynamical description of the systems (i. e. equations of motion).
In this regard, the starting point is the {\it a priori} existence
of a parameter $t$, called from now onwards ``parametric time",
which describes the Newtonian concept of time. It is a fundamental
part of a structure of reality constituted by an inert background
in which dynamics takes place, but which, paradoxically, lacks on
its turn of a dynamic character.

It must be emphasized that the equations of physics do not contain
magnitudes in themselves but rather their measurements. It is thus
more in the scope of physics to speak about ``clocks" or
``clock-time" than about time. In more intuitive terms, the
problem can be posed as a question on the dynamical character or
not of the time variable (parametric and deparametrized theories).

Traditionally, quantum physics has stated that the sea of virtual
pairs that pop-up and disappear constantly in empty space, {\it i.
e.} the quantum vacuum, has infinite energy density as follows from
the simple application of its basic principles. However, there is
now some evidence that it may be finite. In fact, it is not
understood why this density seems to be so small as is shown by its
possible cosmological manifestations. The question is important
since the quantum vacuum fixes the values of some observable
quantities, as the electron charge and the light speed, or gives
rise to observable phenomena such as the Casimir effect or the Lamb
shift.

The plan of this paper is as follows. Section 2 is devoted to
discuss the inclusion of a dynamical time in Hamiltonian formalism.
It is shown in section 3 that a coupling between the background
gravitation that pervades the universe and the quantum vacuum is
unavoidable, and that this determines that atomic clocks must march
differently from the astronomical ones. In section 4, we look for an
observable effect of this coupling, the best candidate being the
so-called Pioneer anomaly. Section 5 shows that the coupling is not
in conflict with astronomical data. Finally in section 6 we state
our conclusions.

\section{Time and clocks}

In classical dynamics the physical time appears as a non-dynamical
variable that allows the expression of the action integral with
the form $S=\int L {\rm d} t$. As a consequence of this
non-dynamical character, does not exist a canonical momentum
conjugate to $t$. From the Hamiltonian \begin{equation}
H(p,q)=\sum p\,\dot{q}-L\,,\label{30}\end{equation} the equations
of motion in the standard form are $\dot{q}=\partial H/\partial
p$, $\dot{p}=-\partial H/\partial q$. To translate this formal
machinery to a scheme in which time acquires the character of a
dynamical variable, there exists a canonical approach which allows
an interpretation in terms of ``clock-time" with its specific
dynamical variables \cite{Tie02,Bar98,Han76}. Instead of the
deparametrized action $S=\int [p\,\dot{q}-H(p,q)]{\rm d} t$, we
can use the alternative form
\begin{equation} S=\int\{\Pi
(t)\dot{\sigma}(t)+p\,(t)\dot{q}(t)-u(t)[ \Pi(t
)+H(p\,(t),q(t))]\}{\rm d} t \,,\label{40}\end{equation}
 (overdot means derivation with respect to the parameter $t$) where
\mbox{$H(p\,(t),q(t))$} has the same functional form as the
Hamiltonian in (\ref{30}), $t$ is  introduced in such a way that the
theory becomes invariant with respect to reparametrizations and
$\sigma (t),\,\Pi (t)$ are conjugate dynamical variables that
describe the clock. Note that $\Pi _u$, the momentum conjugate to
$u(t)$, weakly vanishes.

The corresponding Hamiltonian writes \begin{equation} \hat{H}=u[\Pi
+H(p,q)]+\lambda \,\Pi _u\,,\label{45}\end{equation} where $\lambda$
is a Lagrangian multiplier. The stability of the weak condition $\Pi
_u=0$ implies $\Pi +H(p,q)=0$. Both are first-class constraints ({i.
e.} symmetries). The transformations induced by $\Pi _u$ allow then
to interpret $u(t)$ as an arbitrary function which can be considered
as non-dynamical. The extended Hamiltonian is then
\begin{equation} H^E =u[\Pi +H(p,q)]\,,\label{50}\end{equation}
being hence singular as far as it is proportional to the scalar
constraint. Simple algebra allows to verify that $\Pi +H(p,q)$ is
the reparametrization generator, as requested by the invariance
properties of the action (\ref{40}). The equations of motion are
then \begin{equation} \dot{q}={\partial H^E\over
\partial p}=u{\partial H\over \partial p}\,;\quad  \dot{p}=-{\partial H^E\over
\partial q}=-u{\partial H\over \partial q}\,;\quad u=\dot{\sigma}\,;\quad
\dot{\Pi}=-\dot{H}=0\,.\label{60}\end{equation} From these
equations it follows that
\begin{equation} {{\rm d} q\over {\rm d} \sigma}={\partial H\over
\partial p}\,;\quad {{\rm d} p\over {\rm d} \sigma}=-{\partial H\over
\partial q}\,;\quad u={{\rm d} \sigma \over {\rm d} t}\,;\quad {{\rm d}
H\over {\rm d} \sigma}=0\,. \label{70}\end{equation} The first two
are the canonical equations of motion with $\sigma$ as the time
variable. The third one expresses what we call the ``march" of the
clock $u$ with respect to the parameter $t$. This theory being
invariant under reparametrization, we may fix the gauge by the
condition $\sigma =t$ ({\it i. e.} $u=1$), so that we recover the
ordinary canonical formalism with $t$ being the Newtonian time. We
see that with this choice $\sigma$ is the time measured by an ideal
clock, defined as one which can be made to run with the Newtonian
time. As $\sigma$ and $\Pi$ (which is weakly equal to $-H$) are
canonically conjugate variables, the fourth Heisenberg relation
(involving now a dynamical time variable and the energy) acquires
clearly a dynamical meaning. Notice the close analogy between the
third equation (\ref{70}) with the very concept of proper time in
general relativity ${\rm d} \tau =\sqrt{g_{00}}\,{\rm d} t$, where
$\sqrt{g_{00}}$ is the march of a proper clock with respect to the
parametric time. It must be emphasized that the ordinary
deparametrized dynamics is a particular gauge of this scheme. The
extension of this formalism to more complex functional dependence on
the time variables gives essentially the same result, although at
the expense of some unnecessary complexity \cite{Tie02}. Note that
this formulation gives a dynamical basis for the fourth Heisenberg
relation because the energy and the time $\sigma$ are clearly
conjugate dynamical variables with Poisson bracket $[H, \sigma ]=1$.
This precision will be important later.

 As a matter of fact, no criterion exists, other than an arbitrary
choice, to fix the march of a real clock to that of parametric time,
being this one essentially unobservable by its own definition, in a
parametric invariant description. The Hamiltonian (\ref{45}) is the
sum of two terms, describing, respectively, the physical system and
the clock. {\it The equation of motion of the second  term, $H_{\rm
clock}=u\Pi + \lambda \Pi _u$, is precisely that of a clock $u={\rm
d}\sigma /{\rm d}t$}. The situation has to be understood as the
arbitrary definition of standard clock as one that verifies the
relationship $u={\rm d} \sigma _0/{\rm d} t=1$, denoting the time of
the standard clock as $\sigma_0$. Notice that the definition of a
standard clock refers precisely to its march. No change of units is
involved as it happens when scale transformations are present
\cite{Can77}.

The simplest example of a theory with a clock is given by the free
particle in special relativity, described by the action
$$S=-mc\int {\rm d}s\equiv -mc\int \sqrt{\dot{x}_0^2-\dot{x}_i^2}\,{\rm
d}t\,,$$ which is parametric invariant by construction. The
corresponding scalar constraint leads us to the on-shell condition
$p_0^2-p_i^2=m^2c^2$.

The extended hamiltonian analogous to (\ref{50}) reads
$$H^E =u[p_0+H]\,,\qquad \left(H=\mp \sqrt{p_i^2+m^2c^2}\right)\,,$$
showing the role as a clock played by the dynamical variable $x_0$.
The result is essentially the same in the presence of a
non-euclidian metric, both are examples of a class of theories
having a clock as well as parametric invariance. The
Einstein-Hilbert action for pure gravity, quite on the contrary, is
a parametric invariant theory without any clock as a dynamical
variable. This is a situation that is by no means obvious to
understand.

As long as the observations make use of only the standard clock, the
scheme is nothing else than the Hamiltonian equations. This may not
occur, however, if there is another clock with a different march. In
the latter case, the motion equations are (\ref{70}), but with
$\sigma _0$ instead of $t$
\begin{equation} {{\rm d} q\over {\rm d} \sigma}={\partial H\over
\partial p}\,;\qquad {{\rm d} p\over {\rm d} \sigma}=-{\partial H\over
\partial q}\,;\qquad {{\rm d} \sigma \over {\rm d} \sigma _0}=u\,, \label{80}\end{equation}
which describe the physics of a system in operationally realistic
terms. This means that they do not refer to any unobservable
parametric time but to $\sigma _0$ and $\sigma$, which are times
really observed by real clocks. The novelty is here the presence
of the third equation (\ref{80}), which is the dynamic equation of
the second clock with respect to the standard one.

It must be underscored that there is no criterion to determine the
march $u$, other that to refer to the internal properties of the
clock or the pure observation, particularly if they are based on
different phenomena. Physics has accepted traditionally without any
discussion the implicit ``principle" that all kinds of clocks have
the same march and measure the same time. However, {\it if two
clocks are based on different phenomena, they are not necessarily
equivalent, specially in the case where the parametric invariance is
broken}, in the sense that the march ${\rm d} \sigma /{\rm d} \sigma
_0$ may not be a constant (if it were, the two clocks would measure
the same time with different units). In the following, we will use
as standard clock-time $t_{\rm atomic}$, {\it i. e.} the time
measured by the atomic clocks

\section{Coupling between the background gravitation and the
quantum vacuum}

 Following the same
 scheme as in section 2, we assume here a classical parametric invariant
 description, which allows us to choose $t$ as a non-relativistic
 Newtonian time. Let us consider the background gravitational
potential that pervades the universe and let us do it from a
phenomenological viewpoint. The potential near Earth can be written
approximately as $\Psi _{\rm E}({\bf r},t)=\Psi _{\rm loc}({\bf
r})+\Psi (t)$ (here $\Psi=\Phi /c^2$, $\Phi$ being the dimensional
 Newtonian potential).  The first term $\Psi _{\rm loc}$ is the
part of the local inhomogeneities, as the Sun and the Milky Way,
which are not expanding so that it is time independent. The second
$\Psi(t)$ is due to all the mass-energy in the universe assuming
that it is uniformly distributed. Contrary to the first, it depends
on time because of the expansion. The former has a nonvanishing
gradient but is small, the latter is larger but its gradient
vanishes.  In the following $\Psi (t)$ will be called the background
potential of the entire universe. Since the gravity is weak and the
geometry of the universe has approximately flat space sections, we
take the Newtonian approximation.

 Because gravitation is a long range universal interaction that
 affects to all the matter and energy in the universe, there must
be necessarily a coupling between the background potential $\Psi
(t)$ and the quantum vacuum. Consequently, the existence must be
admitted of some kind of adiabatic progressive modification of the
structure of the quantum vacuum in the expanding universe. The
analysis of the previous section shows that the dynamical time and
the energy can be defined as conjugate canonical variables, what
gives a dynamical basis to the fourth Heisenberg relation $\Delta
E\cdot \Delta \tau\approx \hbar$. Let us consider now the sea of
virtual pairs in the quantum vacuum, with their charges and spins,
assuming that its energy density is finite. On the average and
phenomenologically, a virtual pair created with energy $E$
 lives during a time $\tau _0= \hbar /E$, according to the
fourth Heisenberg relation. This has an important consequence: the
optical density of empty space must depend on the gravitational
field. Indeed, at a spacetime point with
 gravitational potential  $\Psi ({\bf r},t)$  , the pairs have an extra
 energy $E\Psi$, so that their lifetime must be
$\tau _\Psi =\hbar/(E+E\Psi)= \tau _0/(1+\Psi).$

The conclusion seems clear \cite{Ran03a,Ran03b}: the number
density of pairs $\cal N$ depends on $\Psi$ as \mbox{${\cal
N}_\Psi ={\cal N}_0/ (1+\Psi)$}. If $\Psi$ decreases, the quantum
vacuum becomes denser, since the density of charges and spins
becomes higher; if $\Psi$ increases, it becomes thinner.
Consequently, the gravity created by mass or energy thickens the
quantum vacuum, while the gravity created by the cosmological
constant or the dark energy attenuates it. This is important,
since the quantum vacuum plays a decisive role to renormalize the
naked or bare values of some quantities to their observed values,
as is the case of the electron charge and the light speed.
 It must be stressed that this  is not an
{\em ad hoc} hypothesis, but a necessary consequence of the fourth
Heisenberg relation and the universality of gravitation. The
effect of the time independent $\Psi_{\rm loc}$ is neglected here
because, as will be shown, the potential acts in this model
through its time derivative.

We accept then the following phenomenological hypothesis: the
quantum vacuum can be considered as a substratum, a transparent
optical medium characterized by a relative permittivity
$\varepsilon _{\rm r}(\Psi)$ and a relative permeability $\mu
_{\rm r}(\Psi)$, which are decreasing functions of $\Psi (t)$. As
$\Psi(t)$ increases, the optical density of the medium decreases
(since there are less charges and spins) and {\it vice versa}.
Therefore, the permittivity and the permeability of empty space
can be written as $\varepsilon=\varepsilon _{\rm r} \varepsilon
_0$ and $\mu =\mu _{\rm r}\mu _0$, where the first factors express
the effect of the gravitational potential, {\it i. e.}, the
thickening or thinning of empty space.   We can write, at first
order in the variation of $\Psi (t)$,
\begin{equation} \varepsilon _{\rm r}(t)=1-\beta [\Psi (t)-\Psi
(t_0) ],\qquad \mu _{\rm r}(t)=1-\gamma [\Psi (t)-\Psi (t_0)]\,,
\label{90}
\end{equation}
 where $\Psi (t_0)$ is
the reference potential at present time $t_0$
 and $\beta$ and $\gamma$ are certain coefficients, necessarily
 positive since the quantum vacuum must be dielectric but paramagnetic
(its effect on the magnetic field is due to the magnetic moments of
the virtual pairs). The results of this paper will depend only on
the semisum $\eta =(\beta +\gamma)/2$. Not surprisingly, it turns
out that $\Psi (t)$ varies adiabatically (because of the expansion)
and that its time derivative at present time
 $\dot{\Psi}_0=\dot{\Psi}(t_0)$ is positive and very small, of the order of
 $H_0$ as we will see later. It must be underscored that eqs. (\ref{90}) express
 a modification of the structure of the quantum vacuum as an effect
 of its coupling with the background gravitation in the expanding
 universe.

 It is easy to show that if
$\varepsilon _{\rm r}$ and $\mu _{\rm r}$ decrease adiabatically
as (\ref{90}) (and the optical density of empty space, therefore),
 the frequency and the speed of an
electromagnetic wave increase adiabatically as
\begin{equation} \dot{\nu}/ \nu=\dot{c}/ c=-(\dot{\varepsilon}/\varepsilon
+\dot{\mu}/\mu)/2= \eta\dot{\Psi}_0\,,\label{100}\end{equation}
or, equivalently,
\begin{equation}  \nu =\nu _0[1+\eta \dot{\Psi}_0(t-t_0)]\,,\qquad
c =c_0[1+\eta \dot{\Psi}_0(t-t_0)]\,. \label{105}\end{equation} The
proof is very simple \cite{Ran03a,Ran04}: just write the Maxwell's
equations with $\varepsilon$ and $\mu$ decreasing adiabatically with
time as in (\ref{90}) to find
 that eqs. (\ref{100}) and (\ref{105}) are satisfied at first order in $\dot{\Psi}_0$.
If $\varepsilon ,\,\mu$ vary with time as an effect of the change
of the densities of virtual charges and spins, the speed of light
changes also. This is not dissimilar to what happens to light in a
medium, say diamond, where speed is different from $c$ because the
quantum effects of the lattice add to those of empty space.

As a consequence, the quantum vacuum can be characterized by a
refractive index $n(t)= 1+\eta \dot{\Psi}_0(t-t_0)$ depending on the
Newtonian time. We can interpret this result in two ways: (i) the
first and obvious one is that light accelerates with the Newtonian
time; however, attention must be paid to the fact that this
statement depends on the particular choice of the Newtonian clock,
as was shown before. (ii) Nevertheless, since the dynamical equation
of a clock is its march, it suffices to take a clock with march
relative to the Newtonian one equal to the refractive index
\begin{equation} {{\rm d}\sigma \over {\rm
dt}}=n(t)\,,\label{94}\end{equation} for the frequencies and the
speed of light to be constant (\ref{105}). It is clear that such a
clock is precisely an atomic clock since the periods of an
electromagnetic wave are its basic units.

As we said before, there is no clock in Einstein-Hilbert formulation
of General Relativity. However, the constancy of the light speed
provides us with a dynamical time: the proper time.  In fact the
geometric structure of space-time induces a relative permittivity
$\varepsilon _{\rm r}$ and permeability $\mu _{\rm r}$ of empty
space different from 1, their common value being $\varepsilon _{\rm
r}=\mu _{\rm r}=(g_{00})^{1/2}$ \cite{Lan75}. As a consequence of
diff-invariance, the use of the proper time defined as ${\rm d}\tau
=\sqrt{g_{00}}\,{\rm d}t$ restores the constancy of the light speed.

Being a quantum effect, the coupling to the quantum vacuum breaks
diff-invariance and cannot, therefore, be included in the definition
of proper time. Consequently, it is an alien element that must be
added to general relativity.

Hence, three times must be considered. They are

(i) the parametric time $t$, namely the ephemerids time (which is a
coordinate time), used to calculate the trajectory of the
spaceships.

(ii) The proper time $\tau$ of General Relativity.

(iii) The time of the atomic clocks $t_{\rm atomic}$, defined as
\begin{equation}{\rm d}t_{\rm atomic}=[1+\eta (\Psi (t)-\Psi (t_0)]\,{\rm
d}\tau =[1+\Psi_{\rm loc}+\eta (\Psi (t)-\Psi (t_0)]\,{\rm
d}t\,.\label{A20}\end{equation} We will omit in the following the
term $\Psi_{\rm loc}$ because it is constant and we will be
concerned with the time variation of the potential.

 It is usually assumed that the two dynamical times $\tau$ and
$t_{\rm atomic}$ are in fact the same one. However, this is no
longer true if $\eta \neq 0$.

\section{Looking for the effect}

\subsection{The Pioneer anomaly} Anderson {\it et al.} reported in 1998 a
curious anomalous effect \cite{And98,And02}. It consists in an
 adiabatic frequency blue drift of the two-way radio signals
from the Pioneer 10/11 (launched in 1972 and 1973), manifest in a
residual Doppler shift, which increases linearly with time as
\begin{equation} \dot{\nu}/\nu =2a_{\rm t}\,,\quad \mbox{or}\quad
\nu =\nu_0\,[1+2a_{\rm t}(t-t_0)]\,,\label{10}\end{equation} where
$t_0$ is the launch time, $a_{\rm t}= (1.27\pm 0.19) H_0$ and $H_0$
is the Hubble constant (overdot means time derivative). More than 30
years afterwards, the phenomenon is still unexplained. Since it was
detected as a  Doppler shift that does not correspond to any known
motion of the ships, its simplest interpretation is that there is an
anomalous constant acceleration towards the Sun. However, this  is
not acceptable since it would conflict with the well known ephemeris
of the planets and with the equivalence principle.

Anderson attempted a second interpretation: $a_{\rm t}$ would be ``a
clock acceleration", expressing a kind of inhomogeneity of time.
They imagined it in an intuitive and phenomenological way, without
any theoretical foundation, saying that, in order to fit the
trajectory, ``we were motivated to try to think of any {\it (purely
phenomenological) `time' distortions that might fortuitously} fit
the Pioneer results" (our emphasis, ref. 2, section XI). They obtain
in this way the best fits to the trajectory (using the adjective
``fascinating"). In one of them, which they call ``Quadratic time
augmentation model", they add to the TAI-ET (International Atomic
Time-Ephemeris Time) transformation the following distortion of ET
\begin{equation} \mbox{ET} \rightarrow \mbox{ET}^\prime=\mbox{ET} +{1\over
2}\,a_{\rm ET}\,\mbox{ET}^2\,.\label{20}\end{equation} This means to
take a new time ET$^\prime$ that is a quadratic function of ET and
which, therefore, accelerates with respect to ET. Note that ET is a
parametric coordinate time while TAI is a dynamical time. But they
gave up the idea because of the lack of any theoretical foundation
and contradictions with the determination process of the orbits.
They were not on the back track, however, as will be seen here.

In 1998, after many unsuccessful efforts to account for it, the
discoverers suggested ``that the origin of the anomalous signal is
new physics" \cite{And98}. In other words, to try non-standard ideas
could be a good strategy to solve the riddle, specially if they are
not in conflict with current physics, as is the case with the
proposals of this work.

\subsection{A problem of dynamics of time}
From now on, the ephemeris time will be called ``astronomical time"
and noted $t_{\rm astr}$. It turns out, as will be shown, that the
anomaly has the same observational signature as an acceleration of
the marches of atomic and the astronomical clocks with respect to
one another. In other words, as the deceleration of the astronomical
clocks with respect to the atomic clocks (or, equivalently, to the
acceleration of the latter with respect to the former).   This might
seem surprising since it is always assumed as a matter of fact that
both types of clocks measure the same time. This is not necessarily
true, however, since they are based on different physical laws. The
importance of this point must be stressed. The {\it astronomical
clock-time}, say $t_{\rm astr}$, is defined by the trajectories of
the planets and other celestial bodies. It is measured with
classical and gravitational clocks, the solar system for instance.
On the other hand, the {\it atomic clock-time}, say $t_{\rm
atomic}$, is founded on the oscillations of atomic systems. It is
measured using quantum and electromagnetic systems as clocks, in
particular the oscillations of atoms or masers. Note that, contrary
to the concept of time, which is subtle and difficult, the idea of
``clock"\ is clearly defined from the operational point of view.
This is done by means of certain dynamical systems, the clocks, in
such a way that the time measured by each one is a dynamical
variable, the angular position of a pointer for instance. The
measured time could be different from one clock to the other since,
at least in principle, they could tick at different rates, even at
the same place and having the same velocity. Indeed, eq. (\ref{10})
can be understood as a progressive decrease of the period of the
radio signal, so that the basic unit of the atomic clock-time would
be decreasing with respect to the astronomical time of the orbit.

It is clear that the two clock-times are very close at least but,
since we lack a unified theory of gravitation and quantum physics,
the assumption that they are exactly the same $t_{\rm astr}
=t_{\rm atomic}$ must not be taken for granted. In the explanation
proposed here, the two times are different because of the
previously discussed coupling between the quantum vacuum and the
background gravitation, in such a way that $a_{\rm t} ={\rm d}
^2t_{\rm atomic}/ {\rm d} t_{\rm astr} ^2$, which gives a precise
meaning to what Anderson {\it et al.} called ``clock
acceleration": {\it it is the acceleration of the time of the
quantum atomic clocks with respect to the astronomical time, this
one determined by purely gravitational and classical phenomena}.
In such a way, the Pioneer becomes a two-clock system: the
astronomical clock of the orbit and the atomic clock that measures
the time of the devices used to track the trajectory, which are
based on quantum physics. The ideas expounded here follow from the
confluence of two research lines, one on the Pioneer Anomaly
itself \cite{Ran03a,Ran03b,Ran04}, the other on the dynamics of
time \cite{Tie02,Bar98}.

\subsection{The deceleration of the astronomical clocks}
As already said, eq. (\ref{10}) suggests that the periods of the
microwaves from the Pioneer are decreasing with respect to the
time used to define the orbit; in other words, that the atomic
clocks accelerates with respect to the astronomical clock.  Eqs.
(\ref{100})-(\ref{105}) show that something similar happens as a
consequence of the coupling between quantum vacuum and
gravitation, if $\dot{\Psi}_0>0$. More precisely that the
frequencies and the light speed increase if defined with respect
to $t_{\rm astr}$. However, using instead the time $t _{\rm
atomic}$, defined through
\begin{equation} {\rm d} t _{\rm atomic}=[1+\eta \dot{\Psi}_0\,(t_{\rm astr} -t_{\rm astr,\,0})]\,{\rm d}
t_{\rm astr} \,,\;\; \mbox{so that}\;\; \left.{{\rm d} ^2t_{\rm
atomic}\over {\rm d} t_{\rm astr} ^2}\,\right|_0=\eta \dot{\Psi}_0
=a_{\rm t} \label{110} \end{equation} there is no anomaly, since
both the frequency and the light speed are then constant with this
time. It is clear that $t_{\rm atomic}$ is the time measured by the
atomic clocks, since the periods of the atomic oscillations, which
are decreasing if measured with the astronomical time
 (as shown by (\ref{100}) where $t=t_{\rm astr}$), are
obviously constant with respect to $t_{\rm atomic}$ itself, in fact
they are its basic units. The meaning of $a_{\rm t}$ is neat also:
it is the acceleration of the atomic clock-time with respect to the
astronomical clock-time (note that it would be zero without the
quantum vacuum, since $\eta=0$ then). Alternatively, it could be
said that $-a_{\rm t}$ is the deceleration of the march of the
astronomical clock-time with respect to the atomic clock-time. It
follows then from (\ref{110}) and (\ref{10}) that what Anderson {\it
et al.} observed could well be the march of the atomic clocks that
tracked the ship with respect to the astronomical clock of its orbit
$u = 1+\eta \dot{\Psi}_0\,(t_{\rm astr} -t_{\rm astr,\,0})$. After
synchronizing the two so that they give the same time now, $t_{\rm
astr,\,0}=t _{\rm atomic,\,0}=0$, eqs. (\ref{110}) can be written in
the two equivalent ways
\begin{equation}  t_{\rm atomic}=t_{\rm astr} +{1\over 2}\, \eta \dot{\Psi}_0\,t_{\rm astr}
^2,\quad t_{\rm astr}=t_{\rm atomic} -{1\over 2}\, \eta
\dot{\Psi}_0\,t_{\rm atomic} ^2\,, \label{120}\end{equation} near
present time. The first coincides with eq. (\ref{20}) if $t_{\rm
astr}=\mbox{ET}$ and $t _{\rm atomic}=\mbox{ET}^\prime$. The
importance of this equation must be stressed since it explains why
Anderson {\it et al.} obtained good fits by distorting
phenomenologically the time.

All this shows (eqs. $\!\!\!$(\ref{100})-(\ref{105})) that the
effect of the quantum vacuum would be to accelerate adiabatically
the light and to increase progressively the frequencies if they are
measured with the astronomical time $t_{\rm astr}$. However they are
constant if measured with the atomic time $t _{\rm atomic}$.
Synchronizing these two times and taking the international second as
their common basic unit at present time, then ${\rm d} t_{\rm
astr}={\rm d} t_{\rm atomic} $ (ref. 5). This means that we can keep
the same symbol for the two derivatives ${\rm d} \Psi/{\rm d} t_{\rm
astr}={\rm d} \Psi /{\rm d} t_{\rm atomic}=\dot{\Psi}_0$ at present
time.

If the march $u={\rm d}t_{\rm atomic}/{\rm d}t_{\rm astr}=1+\eta
\dot{\Psi}_0\,(t_{\rm astr} -t_{\rm astr,\,0})$ is not constant, the
speed measured with Doppler effect and devices sensible to the
quantum time, say $v_{\rm atomic}={\rm d} \ell /{\rm d} t_{\rm
atomic}$, would be different from the astronomical speed $v_{\rm
astr}={\rm d} \ell /{\rm d} t_{\rm astr}$. In the case of the
Pioneer anomaly, $u>1$ (because $t_{\rm astr}>t_{\rm 0,\,astr}$), so
that
\begin{equation} v_{\rm atomic}=v_{\rm astr}\,{{\rm d} t / {\rm d}
t_{\rm atomic}}= v_{\rm astr}[1-a(t_{\rm astr}-t_{\rm
astr,\,0})]<v_{\rm astr}\,.\label{60a}\end{equation} Since the
gravitation theory gives $v_{\rm astr}$ while the observers measure
$v_{\rm atomic}$, {\it there must be a discrepancy between theory
and observation}. An apparent but unreal violation of standard
gravity would be detected.

\subsection{Explanation of the anomaly} These arguments give a
compelling explanation of the anomaly as an effect of the
discrepancy between these two times. Let us see why. The frequencies
measured by Anderson {\it et al.} are standard frequencies defined
with respect to atomic time $t_{\rm atomic}$, since they used
devices based on quantum physics. {\it They did not measure
frequencies with respect to the astronomical time}. However, since
(i) the trajectory is determined by standard gravitation theories
with respect to astronomical time and (ii) the observation used
atomic time, they found a discrepancy with their expectations: this
is the Pioneer Anomaly.

To be specific, the discrepancy consists in the following. The
distance travelled  by a ship along a given trajectory can be
expressed in two ways
\begin{equation} d=\int _0^{t_{\rm atomic}}\!\!\!\!\!\!\!\!v_{\rm atomic}(t')
{\rm d}t'=\int _0^{t_{\rm astr}}\!\!\!\!\!v_{\rm astr}(t'){\rm d}
t'\,,\label{180}\end{equation} where $t'=t_{\rm atomic}$ in the
first integral, $t'=t_{\rm astr}$ in the second and $(0,t_{\rm
atomic})$ and $(0,t_{\rm astr})$ are the same time interval
expressed with the two times (because $v_{\rm atomic} =v_{\rm
astr}/u$ and ${\rm d}t_{\rm atomic}=u{\rm d}t_{\rm astr}$). Equation
(\ref{180}) is always valid, if the two times are equal and if they
are different as well. What we are proposing as the solution to the
riddle is that the two times are in fact different. If, however,
they are assumed to be equal $t_{\rm astr}=t_{\rm atomic}$, then the
distance deduced from observations $d_{\rm observ}$ and the expected
distance according to standard gravitation theory $d_{\rm theory}$
up to the same $t_{\rm atomic}$, which are
\begin{eqnarray} d_{\rm observ}&=&\int _0^{t_{\rm atomic}}\!\!\!\!\!\!\!\!\!\!\!\!v_{\rm
atomic}(t')dt'\,,\quad d_{\rm theory}=\int _0^{t_{\rm
atomic}}\!\!\!\!\!\!\!\!\!v_{\rm astr}(t'){\rm d} t'\,,\nonumber
\\ & &\label{190}\end{eqnarray} with $t'=t_{\rm atomic}$ in both integrals, would be different.
Now, if $u>1$ as during the Pioneer flight, the atomic velocity is
smaller so that $d_{\rm observ}<d_{\rm theory}$: it would seem that
the ship runs through a smaller than expected distance. Apparently,
it would lag behind the expected position.

\begin{figure}[h]
\begin{center}
\scalebox{1}{\includegraphics{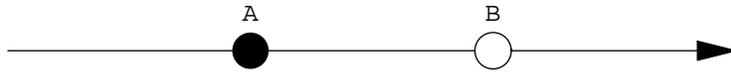}}
\end{center}
\caption{Schematic
representation with arbitrary units of the
 Pioneer anomaly (explanation in the text.)} \label{Fig1}
\label{Figure1}
\end{figure}

All this is explained in Figure \ref{Fig1}, where  the Pioneer
trajectory receding from the Sun is plotted schematically. The
spacecraft moves in the sense of the arrow. The white circle at $B$
is the position of the ship according to standard gravitation, its
real position in fact if this theory is correct. The black circle at
$A$ is the apparent position, deduced from the the ship velocity
after measuring with atomic clocks and devices the Doppler effect on
the frequencies of the signals. If the two times were the same, as
usually assumed, $A$ and $B$ would coincide; if they are not, as in
this model, $A$ and $B$ separate and $A$ would be an apparent
position only.  In the latter case, which one of the pair $A,\,B$
moves faster depends on the value of the relative march of the
atomic clocks with respect to the astronomical clocks $u={\rm d}
t_{\rm atomic}/{\rm d}t_{\rm astr}$.

Since in any time interval during the flight of the ship ${\rm d} t
<{\rm d} t_{\rm atomic}$, the speed measured with Doppler effect and
devices sensible to the quantum time, $v_{\rm atomic}$, must be
smaller than the astronomical speed $v_{\rm astr}$. {\it There would
be an unexplained Doppler residual, easily interpretable  as an
anomalous acceleration towards the Sun.  }Indeed, in the case of one
way signals, $v_{\rm atomic}=v_{\rm astr}{{\rm d} t / {\rm d} t_{\rm
atomic}}= {v_{\rm astr}/ u}=v_{\rm astr}/[1+\eta \,\dot{\Psi}_0(t
-t_0)]$ while for two-way signals
\begin{equation} v_{\rm atomic}=v_{\rm astr}/u^2 =v_{\rm astr}[1-2\eta
\dot{\Psi}_0(t-t_0)] <v_{\rm astr}\,,\label{200}\end{equation} at
first order. In other words, the ship would seem to recede from the
Sun more slowly than expected. There would be an extra blueshift,
since the failure to include the acceleration ${\rm d} ^2t_{\rm
atomic}/{\rm d} t_{\rm astr}^2$ in the analysis mimics a blue
Doppler residual $\dot{\nu}/ \nu= 2\eta\,\dot{\Psi}_0$ or $\nu =\nu
_0[1+2\eta \dot{\Psi}_0(t-t_0)]$ (compare with (\ref{10})). This is
precisely what Anderson {\it et al.} observed and gives the right
result if $\eta \dot{\Psi}_0\simeq 1.3 H_0$.

 The preceding arguments explain the statement made at the
beginning of section 2 that the Pioneer anomaly has the same
observational footprint as an acceleration of the atomic clock-time
with respect to the astronomical time (or as a deceleration of the
latter with respect to the former, what is the same).

For this model to be right, it is necessary that $\dot{\Psi}
_0>0$, i.e. that the present value of the time derivative of the
background potential be positive. It could be argued that this can
be considered in fact a prediction of the model. Alternatively, a
simple argument shows that $\Psi(t)$ is increasing now. The
potential $\Psi$ can be taken to be the sum of two terms, one due
to the matter, either ordinary and dark, and another to the
cosmological constant or the dark energy. The first is negative
and increases with time because the galaxies are separating; the
second is positive and increases with the radius of the universe.
Indeed their values are proportional to $-1/S$ and $+S^2$,
respectively, where $S$ is the scale factor. This can be further
elaborated with a simple model of $\Psi$, in which $\dot{\Psi}_0$
is positive and of order $H_0$ (see \cite{Ran04}).

\section{Agreement with the gravitational redshift
and other observations}The coupling background gravity-quantum
vacuum affects the astronomical time but does not change the atomic
time. For this reason, this model predicts the same standard values
for the frequencies of all the spectral lines because it does not
affect the measurements made with atomic clock-time. Any one of
these lines, for instance the 1,420 MHz line of the hyperfine
structure of Hydrogen, known with an accuracy of $10^{-14}$, has in
this model exactly the same value as in the tables of physical
constants or data. The reason is simple: this frequency is
calculated in atomic theory and measured with devices based on
quantum physics, which use therefore the atomic time $t_{\rm
atomic}$, not the astronomical time $t_{\rm astr}$. These include
lasers, masers, transponders, prisms, diffraction gratings or
spectrographs. What is new in this work is that the frequencies
defined with respect to $t_{\rm astr}$ are different but these are
not usually measured, probably never.

This model obviously complies with the experiments on gravitational
redshift, because they are performed with atomic clocks, which are
exactly the same thing in this work as in standard physics.  In
particular, it agrees with Einstein formula
\begin{equation} \Delta \nu /\nu =-\Delta \Psi _{\rm loc}\,,
\label{210}\end{equation} However, since the standard analysis is
based on the equality of proper and atomic times, it can be
clarifying to compare the two approaches. In this work and taking
into account the local potential by means of ${\rm d}\tau =(1+\Psi
_{\rm loc}){\rm d}t$, one has ${\rm d}t_{\rm atomic}= [1+\eta
\dot{\Psi}_0(\tau -\tau _0)]{\rm d}\tau$ at first order. The
relative difference between the two approaches is therefore of order
$\eta \dot{\Psi}_0\tau_{\rm flight}$, where $\tau _{\rm flight}$ is
the flight time of the light beam. In the most precise observations
by Levine and Vessot, with accuracy $\simeq 2\times 10^{-4}$, this
time is  $\simeq 3.3\times 10^{-2}\mbox{ s}$ \cite{Wil93}. The
condition on the clock acceleration is then $a=\eta \dot{\Psi}_0<
6\times 10^{-3}\mbox{ s}^{-1}$; $a$ is surely much smaller that this
bound, probably of the order of $H_0$, otherwise the effect would
have been detected before. The difference between $\tau$ and $t_{\rm
atomic}$ is thus too small to de detected in experiments of this
kind.

It might be argued that the deceleration of $t_{\rm astr}$ with
respect to $t_{\rm atomic}$ predicted by this work would conflict
with the well-known cartography of the Solar System, particularly
with the observed periods of the planets. It is not so, however. The
dominant potential, as indicated before, is that of the Milky Way,
which can be taken as constant at the scale of the Solar System and
equal to $\simeq 6\times 10^{-7}$. The march is then $u={\rm
d}t_{\rm atomic}/{\rm d}t_{\rm astr}=1-6\times 10^{-7}$, which is
constant. This means that the equations of the Solar System in
Newtonian physics
\begin{equation} \frac{{\rm d} ^2{\bf r}_i}{{\rm d} t_{\rm astr}^2}
=-\sum _{j\ne i}Gm_j\frac{({\bf r}_j-{\bf r}_i)}{|{\bf r}_j-{\bf
r}_i|^3}\,, \label{220}
\end{equation}
are the same with the two times $t_{\rm astr}$ and $t_{\rm atomic}$.
The reason is clear: These equations  are invariant under the
transformation $t_{\rm astr}\rightarrow t_{\rm atomic}=u\, t_{\rm
astr}$, $G\rightarrow G/u^2$ (they would not be invariant if $u$ is
variable). The third Kepler law, for instance, is equally valid with
$t_{\rm astr}$ and $G$ as with $t_{\rm atomic}$ and $G/u^2$. Indeed,
the best value of $G$,  obtained with atomic clocks, must be in fact
$G/u^2$. There can be no conflict with the cartography of the Solar
system.

\section{Summary and conclusions}

Our conclusions can be stated as follows.

(i) The natural framework to describe dynamics is a parametric
invariant formulation including the time as a dynamical object (a
clock-time). The equation of motion of a clock is precisely its
march.

(ii) Because gravitation is long range and universal since it
affects all kinds of mass or energy, a coupling must exist
necessarily between the background gravitation that pervades the
universe and the quantum vacuum. This coupling can be estimated from
the fourth Heisenberg relation and implies a progressive attenuation
of the quantum vacuum in the sense that the refractive index
$n(t_{\rm astr})$ is a decreasing function of the astronomical time.
However $n(t_{\rm atomic})=1$.

(iii) As argued before, the unexplained Pioneer anomaly (\ref{10})
can be understood as the adiabatic decrease of the periods of the
atomic oscillations with respect to the astronomical time. {\it In
other words, the solar system, considered as a clock, would run
progressively slower than the atomic clocks.} This would be an
effect of the interaction between the background gravitational
potential and the quantum vacuum and, therefore, a certain evidence
of the interplay between gravitation and quantum physics. A
consequence of this coupling would be an acceleration of the quantum
clock-time of the atomic clocks $t_{\rm atomic}$ with respect to the
classical astronomical clock-time $t_{\rm astr}$, equal to what
Anderson {\it et al.} called the ``clock acceleration", $a_{\rm
t}={{\rm d} ^2t_{\rm atomic}/ {\rm d} t_{\rm astr}^2}|\,_0>0$ (or,
equivalently, there would be a deceleration of $t_{\rm astr}$ with
respect to $t_{\rm atomic}$). The relation with the coupling
background gravity-quantum vacuum is given as $a_{\rm t}=\eta
\dot{\Psi}_0$, where $\dot{\Psi}_0$ is the present time derivative
of the background potential and $\eta$ a coefficient that refers to
the structure of the vacuum defined in section 3 (after equation
(\ref{90})). In other words, we propose here that what Anderson {\it
et al.} observed is the relative march of the atomic clock-time of
the detectors with respect to the
 astronomical clock-time of the orbit $u={\rm d}t _{\rm atomic}/{\rm
 d}t_{\rm astr}
 =1+\eta \dot{\Psi}_0(t_{\rm astr}-t_{\rm astr,\,0})$ or, more precisely, its square
 $1+2\eta \dot{\Psi}_0(t_{\rm astr}-t_{\rm astr,\,0})$ because their signal was two-way (compare with (\ref{10})).
Although this new idea may seem surprising and strange, it
conflicts with no physical law or principle.  In fact, it could be
rejected only by using a theory embracing gravitation and quantum
physics, which does not exists thus far.

(iv) The model here presented gives a qualitative explanation, at
least, of the Pioneer anomaly, which could well be a manifestation
of the mismatch between these two times, and a quantum gravity
cosmological phenomenon therefore. In order to know whether this
explanation can be also quantitative, it is necessary to estimate
the value of the ``clock acceleration" $a_{\rm t}$. However, this
value depends on a coefficient, here called $\eta$, which cannot be
calculated without a theory of quantum gravity. On the other hand,
the Pioneer anomaly could be considered as a measurement of $a_{\rm
t}$ to be used in the future as a test for quantum gravity. {\it In
any case, this explanation agrees with the experimental values of
the spectral frequencies, the periods of the planets and the
gravitational redshift.}

Two final comments. First, as a consequence of the coupling between
the background gravitation and the quantum vacuum, the light speed
would increase with acceleration $a_\ell =a_{\rm t}c$ if defined or
measured with respect to the astronomical clock-time. However it is
constant if measured with the atomic clock-time. In fact, an atomic
clock is the ``natural clock" to define and measure the light speed,
since its basic unit is the period of the corresponding
electromagnetic wave, so that the speed and the frequencies are then
necessarily constant. This means that $c$ is still a fundamental
constant if measured with atomic time.

Second, since the Pioneer anomaly would be a quantum effect which
causes the light speed and the frequency to increase if defined and
measured with astronomical proper time, it would be alien to general
relativity. It must be stressed also that, if we accept that there
are non-equivalent clocks that accelerate with respect to one
another because of a coupling between gravity and the quantum
vacuum, a new field of unexplored physics opens, which includes the
very idea of universal dimensional constant, in particular.

\section{Acknowledgements.} We are grateful to J. Mart\'{\i}n
and R. Tresguerres for discussions. This work has been partially
supported by a grant of the Spanish Ministerio de Educaci\'on y
Ciencia.

\end{document}